Alternate Science Investigations for the Kepler Spacecraft

White Paper:

# Precision Rotation Periods and Shapes of Near-Earth Asteroids

*31 August 2013*

Martin Elvis[#a], José Luis Galache[ab] and Gareth Williams[ab]

a. Harvard-Smithsonian Center for Astrophysics,
b. IAU Minor Planet Center

# Point of contact. Email: melvis@cfa.harvard.edu; phone: 617-495-7442

## ABSTRACT

We propose to use a modest fraction of the re-purposed *Kepler* mission time and apertures to greatly increase the quantity and quality of our knowledge of near-Earth asteroids (NEAs) rotation and shape. NEAs are important for understanding the origins of the Solar System, for selecting targets for robotic and human visits, and for hazardous object deflection. While NEAs are being discovered at a rate of 1000/year, only a ~75/year have well-measured rotation periods and shapes. Not only can the *Kepler* mission greatly increase the numbers of well-determined NEA rotation periods (to ≥1000 in 5 years), but may do so with order-of-magnitude greater precision than is routinely achieved from the ground. This will enable 3-D tomographic maps to be produced for the ~250 of the brighter NEAs. A multi-year science program would enable improved data quality checks, larger samples and additional types of science. All these numbers are preliminary. We list a number of issues to be resolved before this program can be properly assessed.

## 1. PROPOSED SCIENCE PROJECT

*Kepler* has been used to find the smallest objects outside of our Solar System. We propose to re-purpose a fraction of the telescope's time to study the smallest objects *within* our Solar System: the near-Earth asteroids (NEAs). *Kepler* could both qualitatively and quantitatively improve our knowledge of these multiply important objects.

### 1.1 THE IMPORTANCE OF NEAR-EARTH ASTEROIDS

All the goals of NASA's NEOO program — an effective *planetary defense*, the selection of targets for *human spaceflight* (HSF), and *asteroid retrieval* (ARM) require NEA *characterization*. Knowing whether a potentially hazardous object (PHO) is metallic, stony or carbonaceous, whether it is solid, fractured, or rubble, whether it is tumbling, and what size and shape it is, are all essential inputs to threat mitigation decisions. *Scientifically* NEAs have the potential to address key questions in NASA's roadmap: the formation of the Solar System, the source of the Earth's oceans, and the origins of life (Chyba & Sagan 1992). NEAs may even be a source of novel materials (Elvis and Zeng, 2013).

### 1.1 THE NEAR-EARTH ASTEROID CHARACTERIZATION PROBLEM

The importance of NEAs has led NASA to support extensive surveys to discover NEAs. With ongoing upgrades to both the Catalina Sky Survey[1] and the Palomar Transient Factory[2], and the coming on line of Pan-STARRS-2, it is likely that the large majority of 100-meter class NEAs will have been detected within a decade.

However, NEA characterization is falling well behind discovery, forming a bottleneck for science, HSF, ARM, and for PHOs. At the current rate of ~150 NEA light curves/year it will take over 100 years to obtain compositions, sizes, and shapes just of the ~20,000 NEAs larger than ~100 m diameter. A dedicated program of optical spectroscopy, LINNAEUS[3], on a modest (2-meter class) telescope has been proposed to bring composition determination rates up to match the discovery rate.

Here we propose to use *Kepler* to provide a major upgrade to the NEA period and shape determination program utilizing a modest fraction of the time and apertures. The *Kepler* mission can measure NEA rotation periods with up to an order-of-magnitude greater photometric precision than is achieved from the ground. This will enable a higher fraction to have periods determined, and will give better aspect ratios. For the brighter NEAs 3-D tomographic maps can be produced.

---

1 http://www.lpl.arizona.edu/css/

2 http://ptf.caltech.edu/iptf/

3 The **L**arge **IN**novative **N**ear-earth **A**steroid **E**valuation & **U**nderstanding **S**urvey was proposed to the NASA NEOO program in 2013. PI: Martin Elvis.

## 1.2 LIGHT CURVES: ROTATION AND SHAPE MEASUREMENTS

The basic tool used to measure NEA rotation periods is the light curve, a time-series of the brightness of an object. Figure 1 shows a high quality example. The repetition time provides the NEA rotation period, while the amplitude of the light curve gives a first measure of the aspect ratio of the object. Note that, as many small NEAs are irregular, “potato-shaped” objects (Figure 2), the true period is typically double the repetition time of the light curve. Binary asteroids can also be diagnosed from their light curves (e.g. Belton et al. 1996).

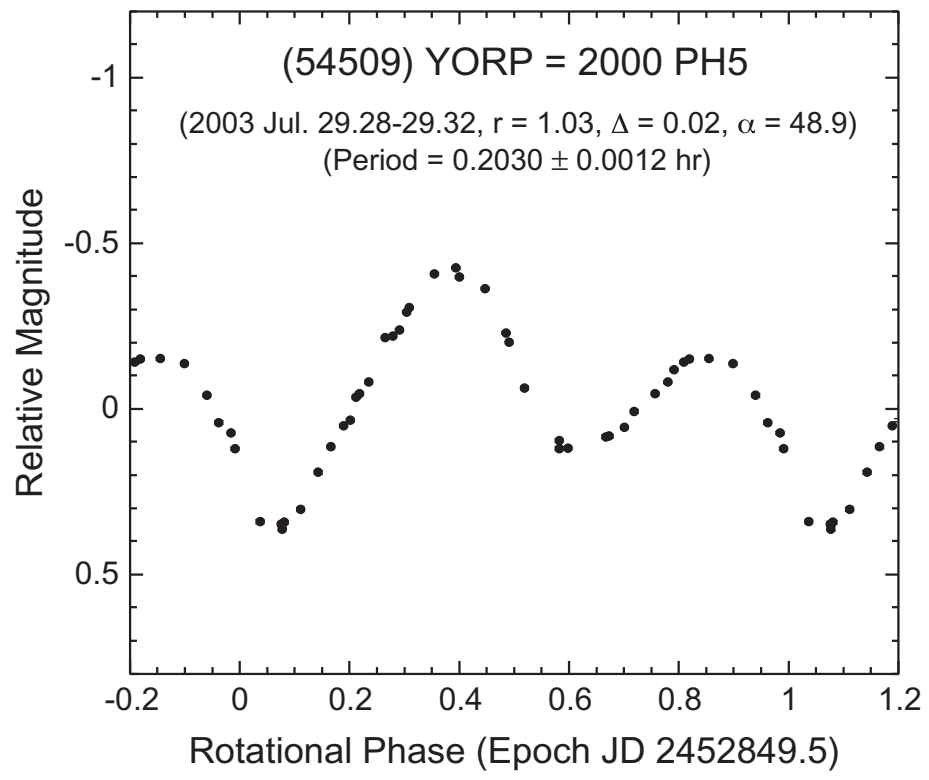

**Figure 1: High quality NEA light curve - for 54509 YORP, 2000 PH5 (Hergenrother and Whiteley 2011).**

The irregular shapes of NEAs show up as non-sinusoidal changes in the light curves. The shape of the light curve depends also on the phase at which it is observed (Harris et al. 1984). Many small NEA are “tumbling”, i.e. rotating about a non-principal axis. Multiple cycles can diagnose a tumbling state as on each rotation a tumbling NEA presents a different face to us, and so a different shaped light curve. Tumbling complicates any spacecraft rendezvous, so just knowing the tumbling rate is valuable.

Tumbling can be used to advantage to determine the 3-dimensional shape of the NEA. Observing many rotations, each one effectively from a different direction in asteroid coordinates, is equivalent to tomography. This allows detailed NEA shapes to be reconstructed. The genetic algorithm SAGE (“Shaping Asteroids with Genetic Evolution”, Bartczak & Marciniak 2012, figure 2) shows how this technique can be quite effective at a few percent photometric accuracy. With *Kepler* it may be possible to reach an order of magnitude better accuracy (Table 1), depending on how well systematic errors can be controlled.

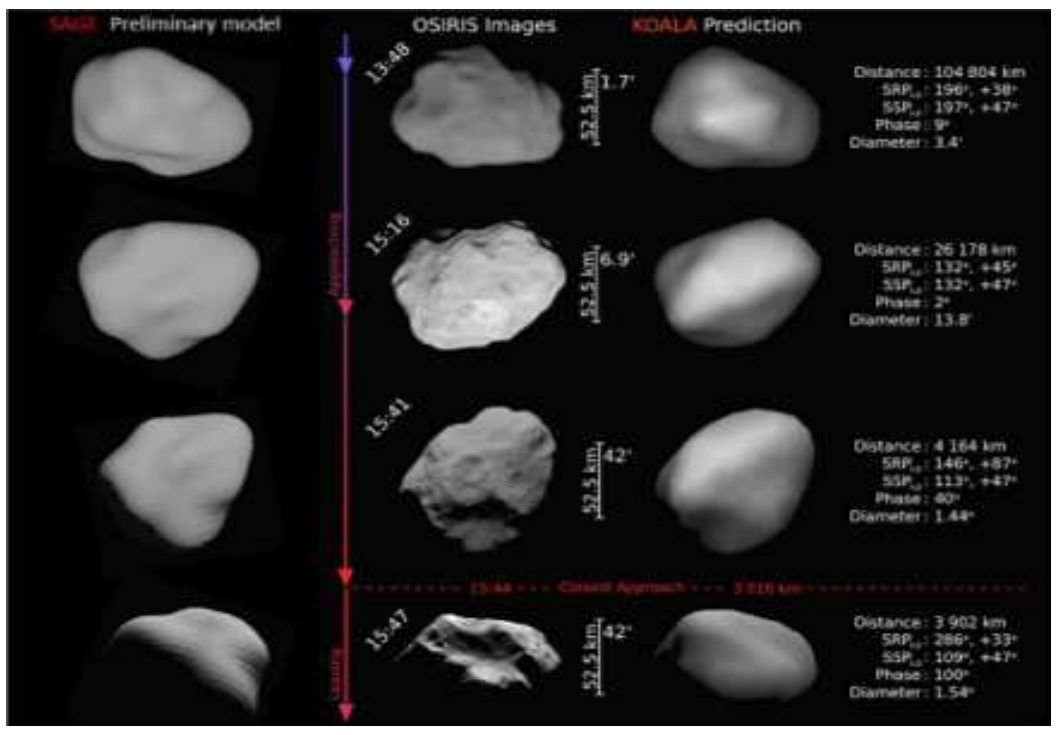

**Figure 2: SAGE model of 21 Lutetia (left) compared with Rosetta fly-by imaging (center) and pre-flyby KOALA model (right) (Barczak and Marciniak, 2012).**

From existing light curves, mostly of Main Belt asteroids, it is apparent that small asteroids, those with H magnitude[4] H < 22, rotate more rapidly than larger asteroids (Figure 3). Large asteroids obey a quite strict limit at 2 hours – almost none spin faster (e.g. Hergenrother and Whitely 2011). Statler et al. (2013) emphasize that the break is remarkably sharp. Faster spinning asteroids will break up if they are held together only by gravity. Such fragmented asteroids are called "rubble piles". A single unfractured rock can spin much faster, although only quite weak forces are needed to allow the observed short periods (Scheeres & Sanchez 2012).

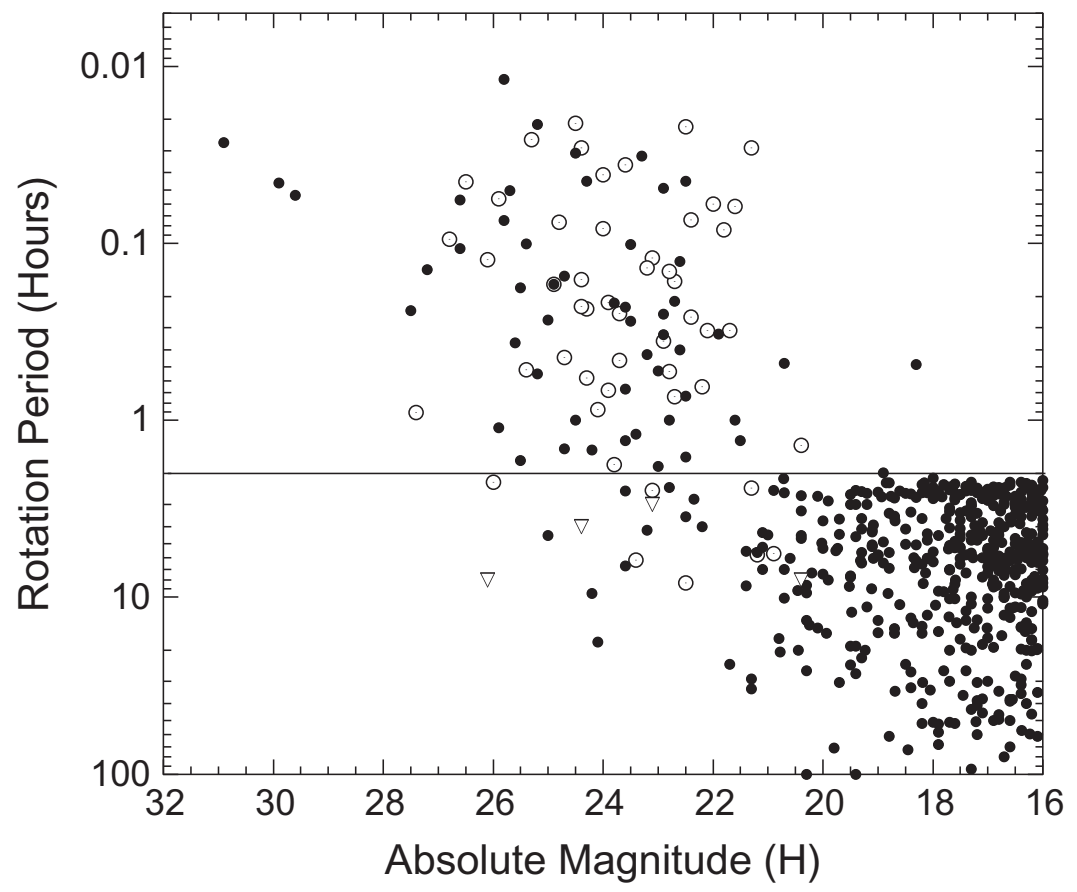

**Figure 3: Rotation period - H magnitude plot. Small asteroids (H > 22) tend to have short rotation periods (< 2 hr), with many in the 0.1 hr range (Hergenrother and Whitely 2011).**

## 1.3 NEA LIGHT CURVES: STATE OF THE ART

Of the 10,000 known NEAs only a few hundred have well-measured rotation periods (IAU Minor Planet Center[5] [MPC] data base, Hergenrother and Whiteley 2011). Only a subset of existing NEA light curves are of sufficient quality to give a high confidence period. A more typical quality light curve is shown in Figure 4. The photometric accuracy of current light curves is typically a few percent. Many observations are taken under non-photometric conditions, through thin cirrus (e.g. Vaduvescu et al. 2013). This moderate signal-to-noise (S/N) makes it statistically hard to distinguish between harmonics, esp. 2:1, and the true periods.

The rotation period uncertainty is quantified via a coded value, U, which takes values from 0 to 3, where 0 means no period can be derived from the data and 3 means a secure result with no ambiguity (Lagerkvist et al., 1989). Less than half (46%) of the 691 NEAs in the MPC database[6] with reported periods have confidently known periods (i.e. U = 3 or 3-). In a uniform survey, Statler et al. (2013) found that only ~25% of NEA light

[4] An asteroid's absolute H magnitude is the visual (V) magnitude an observer would record if the asteroid were placed 1 Astronomical Unit (AU) away, and 1 AU from the Sun and at a zero phase angle (http://neo.jpl.nasa.gov/glossary/h.html). An asteroid with H = 22 has a diameter of 110 – 240 m, depending on the wide range of albedo. Albedo can be delimited if a spectral class is known.

[5] http://www.minorplanetcenter.net

[6] use http://www.minorplanet.info/PHP/lcdbsummaryquery.php to interrogate this database

curves yield a period at a confidence >99%. The remainder may be: (1) spinning too fast (i.e. < integration time); (2) spinning too slowly (i.e. >> observation length); (3) be too spherical to give a measureable amplitude in their light curve. If the amplitude of the light curve is comparable to the error bars on the measurement even a basic aspect ratio cannot be determined.

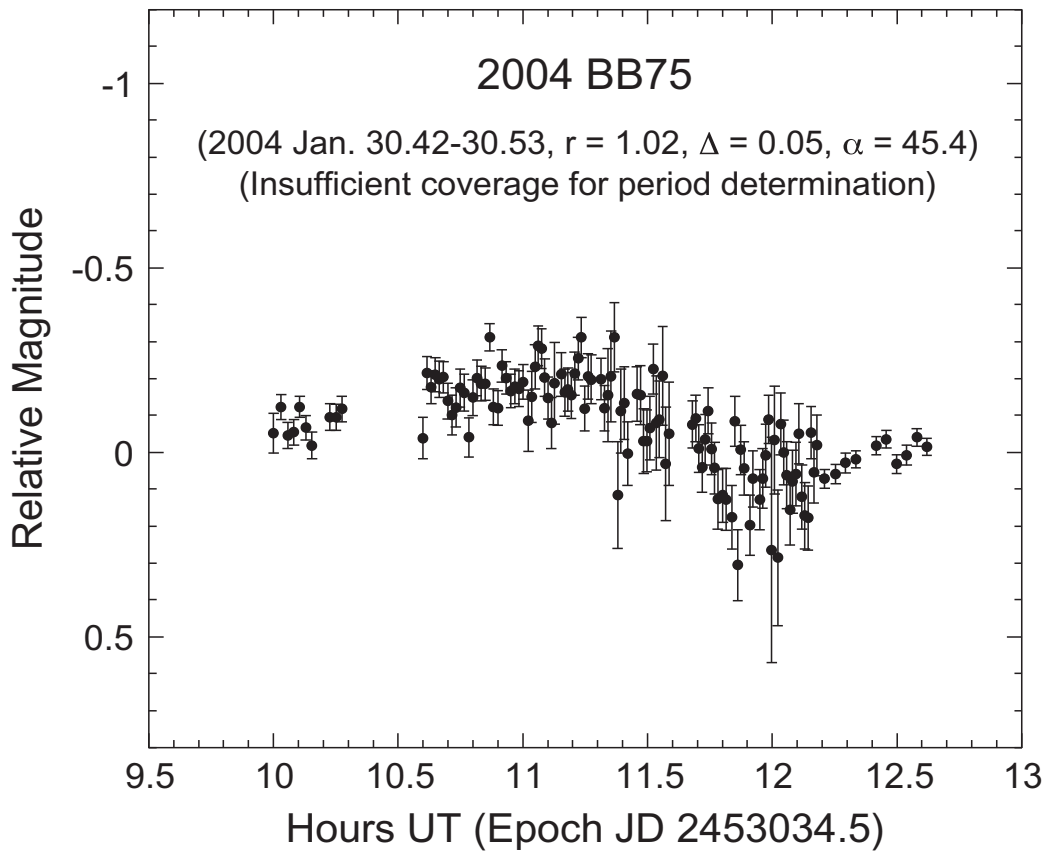

**Figure 4: A more typical quality NEA light curve - 2004 BB75 (Hergenrother & Whiteley 2011).**

At present a variety of relatively small programs and many individuals produce of order 150 NEO light curves annually, of which about half yield well-determined periods, i.e. ~75/year. E.g. Vaduvescu et al. 2013 "EURONEAR" obtained ~100/year over 7 years; Statler et al. 2013 observed 17/year over 5 years. The Palmer Divide Observatory measures ~125 asteroid light curves per year, but most are Main Belt Hungaria objects (Warner 2012a,b, 2013a,b). Almost all are reported quarterly in the Minor Planet Bulletin[7] of the Association of Lunar and Planetary Observers (ALPO), with the data being archived at the MPC.

### 1.4 THE NEED FOR LARGE HOMOGENEOUS UNBIASED SAMPLES OF NEO LIGHT CURVES

The limited number and quality of current NEA light curves restricts the science that can be done with NEAs. As emphasized by Statler et al. (2013), knowing the spin rate distribution as a function of H-magnitude (~size) "*to the accuracy required to constrain the physical properties of NEAs and their dynamical evolution will require larger samples, and homogeneous, unbiased reporting of the data, including accurate errors, for all objects observed, not just those with measured periods*." *Kepler* can address all these issues.

Large unbiased samples will enable correlation of period, elongation (shape), and binarity with other properties beyond size. The orbits, and especially compositions, need to be compared with sizes and shapes. If LINNAEUS, or a comparable program, goes forward then all NEAs with apparitions from ~2014 - ~2017 will have optical Bus-Binzel taxonomic classifications (Bus & Binzel 2002). A large sample ensures that many radar targets will be included, providing a cross-check on size and shape, and also providing masses and mean densities.

---

[7] http://www.minorplanet.info/mpbdownloads.html

### 1.5 RATE OF NEAR-EARTH ASTEROIDS OBSERVABLE WITH KEPLER

To obtain several 100 NEA light curves a year requires reaching to V=18 (Figure 5). Table 2 (Sec.2.3) shows that V = 18 – 18.5 is the approximate limit to which *Kepler* can obtain high quality photometry (by NEA standards). Table 1 gives estimates of the numbers of NEAs within the *Kepler* field-of-regard expected per year in several V and H bins, as well as 1-year and 5-year totals, based on MPC statistics. At V < 18 roughly 300 known and new NEAs cross the *Kepler* field-of-regard (taken to be the Earth-Sun normal ±45 deg) each year. Another 125 have 18 < V < 18.5, for a total of about one a day. At bright magnitudes (V<16) NEAs will appear in the *Kepler* field-of-regard ~55 times/year. The newly discovered NEAs will have a larger fraction at larger H, as a significant fraction of the H < 22 NEAs have already been found (Figure 5).

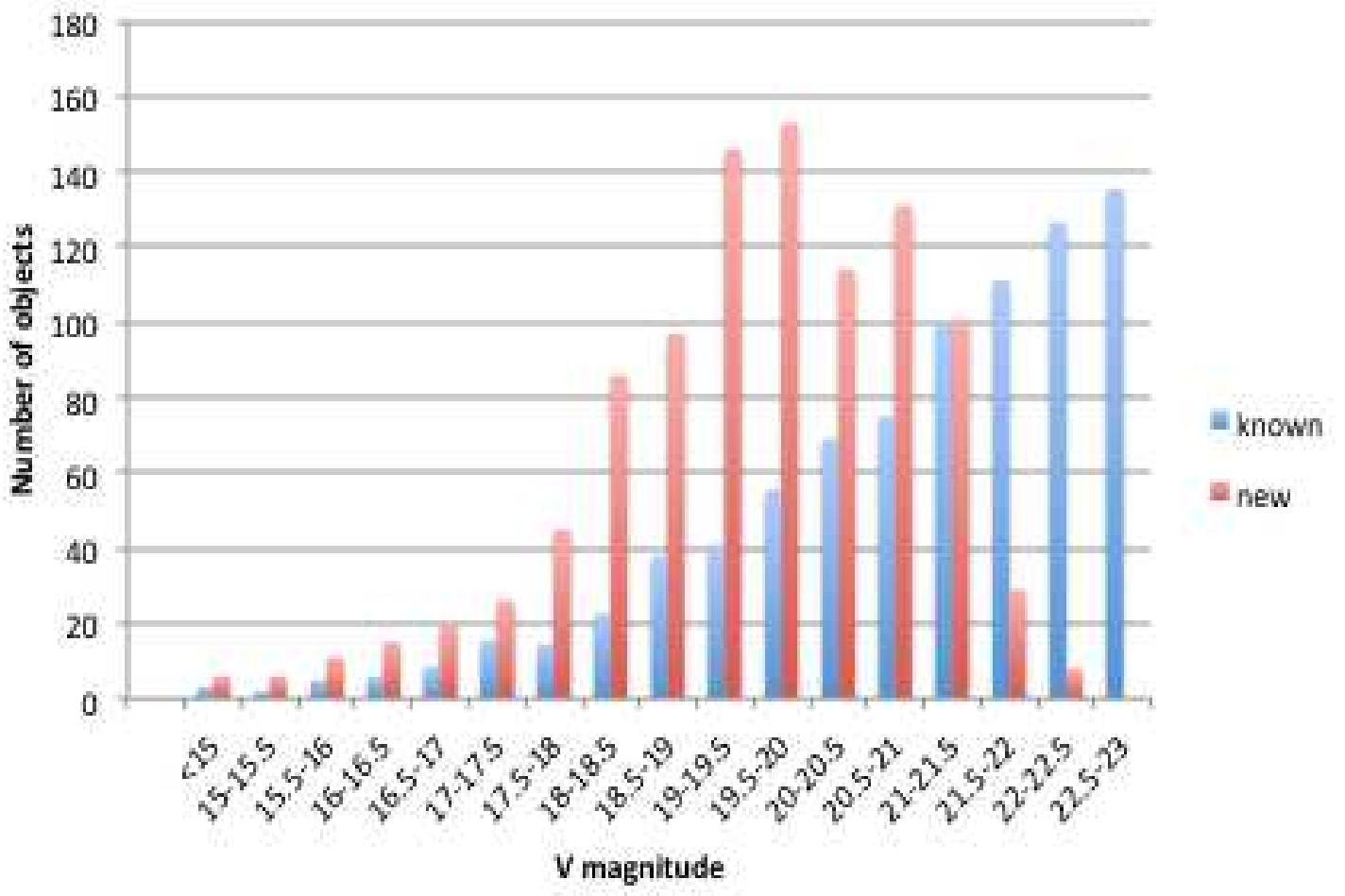

**Figure 5: Predicted V magnitude distribution of NEAs for 2015, divided into new discoveries and recoveries of already known objects (S. Larson, priv. comm.)**

**Table 1: Expected numbers of NEAs, both known and new discoveries by NEA size.**

| NEAs | V < 16 | | V < 18 | | V < 18.5 | | All | |
|---|---|---|---|---|---|---|---|---|
| Size | New | Known | New | Known | New | Known | New | Known |
| H < 22<br>D > 100m (4) | 6 | 41 | 46 | 178 | 76 | 241 | 226 | 604 |
| H > 22<br>D < 240m (4) | 7 | 0.5 | 53 | 5 | 88 | 10 | 274 | 32 |
| H > 28<br>D < 15 m | 0.5 | 0 | 3 | 0 | 3 | 0 | 13 | 0 |
| **1 Year Total** | **13** | **42** | **102** | **191** | 167 | 251 | 513 | 636 |
| **5 Year Total** | **65** | **210** | **510** | **955** | 835 | 1255 | 2565 | 3180 |

# 2. USE OF KEPLER

## 2.1 HOW THE KEPLER SPACECRAFT WILL BE USED

There are two potential modes to carry out this program: active and passive.

### 2.1.1: *Active* Control of the Pointing Direction for a Fraction of the Available Time

For the smaller NEAs (H > 22), which have rotation periods of ~0.1 hours, a maximum of 2.5 hours (25 periods) observing is needed to obtain a period. With photometric errors ~1/3 those of typical NEAs (Table 2) these data will allow more detailed shapes to be determined. Some 60/year will have H ≥ 22 (to V=18) and so will be likely fast rotators.

For the larger NEAs (H < 22), which have rotation periods of 2 – >20 hours (Figure 3), short observations of 1 – 2 minutes spaced over ~1 – a few days at ~1/2 – a few hour intervals will be needed to obtain their periods. Whether this is a feasible observation mode needs to be investigated. There will be ~225/year of these (to V=18).

Bright tumbling NEAs (V ≤ 16) offer the best chance for true tomography as their photometric errors may reach 0.15% (Table 2). A large number of rotation periods, ≥50, needs to be observed for tomography , so a full day of observing is needed for each fast rotator. The ~8/year expected (Table 1) amounts to ~2% of the *Kepler* mission time.

Operational scenarios for the active NEA program could involve observations being either: (1) folded into convenient gaps in other observing programs, which would be good for slowly rotating NEAs, or (2) apportioned dedicated blocks of observing time to avoid splitting up other programs, which would be good for the brighter NEAs.

### 2.1.2: *Passive* use of Pointing, Active use of Apertures.

A fraction of NEAs will pass through the *Kepler* field of view wherever it is pointed. A passive NEA program could use all of this time with only a modest number of apertures being dedicated to following NEAs. This would be comparable to a parallel science observation on the *Hubble Space Telescope*. An advantage of this passive mode is that longer integrations on fainter NEAs could be carried out without significant penalty. If these are larger (H < 22) NEAs then long rotation periods are expected. Our first (crude) estimate is that ~15 NEAs/year will pass through the Kepler field-of-view each year. Hence the Active mode will be more than an order of magnitude more productive.

Proper motions for NEAs are typically ~1"/min (Figure 6, Beeson et al. 2013, Vaduvescu et al. 2013), or 2 arcminutes in a typical minimum observing time of 2 hours. Considering apertures of 4x4 pixels (1 pixel = 3.98 arcsec), a line of ~15 apertures would suffice. If the primary exposure is longer then a longer chain could be defined to follow the NEA across the whole field of view for a many-rotation light curve. These would be particularly useful for slowly rotating NEAs, and for finding binary asteroids.

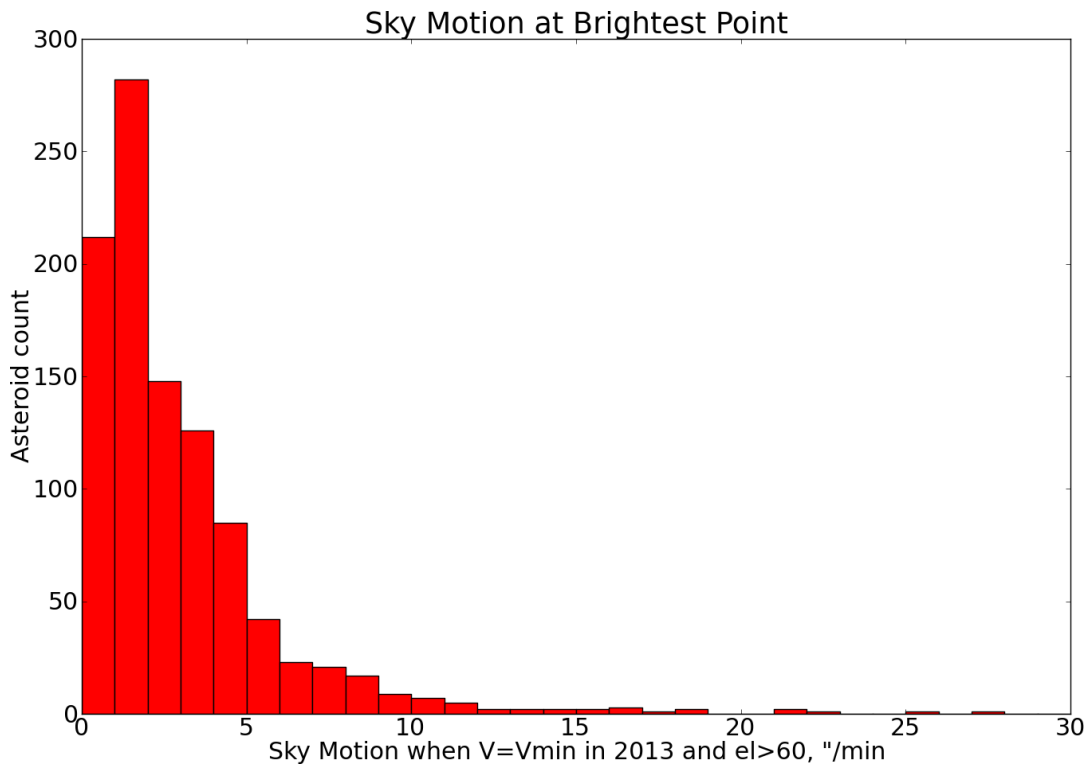

**Figure 6: Proper motion of NEAs (arcseconds/minute) at their brightest V magnitude for their most recent apparition (Beeson et al. 2013).**

## 2.2 HOW THE FOCAL PLANE WILL BE USED

Apertures will be placed contiguously along the path of the NEA on the sky. As noted in Sec.2.1.2 above, proper motions for NEAs are typically ~1"/minute (figure 6), or 2 arcminutes in a typical minimum exposure of 2 hours for which a line of ~15 apertures would suffice. Apertures would avoid stars that are comparable or brighter in magnitude to the target NEA.

## 2.3. PLANNED INTEGRATION TIMES

The Poisson errors from electron counting statistics are given in Table 1, based on the Call for White Papers, for 1 minute integrations for several magnitudes relevant to NEAs . Even down to V=18.5 the error does not exceed 0.6%. These errors are up to an order of magnitude smaller than the 2% - 5% achieved with ground-based photometry and assure that systematic effects will dominate.

**Table 2: Poisson errors for exposures at various V magnitudes with *Kepler.***

| V | t (sec) | N(electrons) | Poisson % error |
|---|---|---|---|
| 12 | 6 | 1.4x10(6)[a] | 0.08 |
| 16 | 60 | 4.4x10(5) | 0.15 |
| 17 | 60 | 1.4x10(5) | 0.27 |
| 18 | 60 | 5.6x10(4) | 0.42 |
| 18.5 | 60 | 3.1x10(4) | 0.57 |

[a.] *from Call for White Papers*

## 2.4 DATA STORAGE NEEDED

TBD (modest)

## 2.5 DATA REDUCTION AND ANALYSIS

Standard tools for analyzing NEA light curves exist. The most used software for lightcurve data reduction and analysis is "Canopus"[8], which provides not only the spin periods derived from the asteroids' photometric observations but also a file containing all the data points and errors in the appropriate ALCDEF (Asteroid Lightcurve Data Exchange Format) form (Warner et al. 2011) that is required for submission to the MPC's light curve database (Warner et al. 2009). PERANSO is also popular (http://www.peranso.com). Standard period-search tools from astronomy packages, e.g. IRAF, are also widely used. The new SALSA package (Statler et al. 2013) will also be investigated. The optimum choice of software to use with *Kepler* data will need to be determined. Custom modifications or front-ends may be needed. Tomography requires the use of more advanced methods using e.g. KOALA (Carry et al., 2012), and SAGE (Bartczak & Marciniak 2012).

[8] http://www.minorplanetobserver.com/MPOSoftware/MPOCanopus.htm

## 2.6 CLASS OF SCIENCE TARGETS

All the NEA targets will be point sources. However they will move across the stellar field at a range of proper motion rates (Figure 6, Beeson et al. 2013).

## 2.7 TARGET DURATIONS

The rotation period-H-mag distribution (Figure 3) shows that small NEAs (H > 22) have periods of 1 - 60 minutes, concentrated at ~10 minutes. Period determination requires ≥5 full periods, so 1 – 2.5 hours should be adequate for small NEAs.

Rotation periods of 10 – 100 hours are common among larger NEAs (H < 22). Within the V < 18 population these comprise the large majority (Table 1), although this is a selection effect. For slow rotators longer elapsed times are needed, ≥2 days for a 10 hour period. But these observations can be built up from 1 – 2 minute snapshots at 0.5 – 1 hour intervals, if telescope slewing constraints can accommodate this sequencing.

For the smaller number of bright NEAs where tomography can be performed the longest data train possible is preferred. Given the *Kepler* field of view and the rate of proper motion of NEAs across the sky this could be a long as 10 days, if spacecraft operations can support this. Such long observations would need to be rare and carefully chosen. A protocol for selecting these targets will need to be developed.

## 2.8 DURATION OF SCIENCE PROGRAM

There is every reason to continue this program for the remainder of the *Kepler* mission:

1. New NEAs are discovered at a rate of ~1000 per year, and the rate is set to increase to ~ 2000 or more a year in 2015 as improved surveys come on line (Catalina Sky Survey, Pan-STARRS 2, Palomar Transient Factory 2). The Catalina Sky Survey alone will discover ~1500 NEAs per year from 2015 onwards (Figure 5, S. Larson, private communication).
2. A longer baseline allows known NEAs a greater chance of returning in a more favorable apparition so that they are brighter.
3. Period changes, if precise measurements are available over years, allow measurement of the YORP effect (e.g. Bottke et al. 2006).
4. Observing non-tumbling NEAs at different axis orientations on different apparitions will aid in reconstructing their 3-dimensional shapes.

## 2.9 SCIENTIFIC IMPACT – DURING AND AFTER PROJECT

This program will improve our knowledge of near-Earth asteroids both quantitatively and qualitatively. Assuming a 5-year program, a total database of >1500 NEO light curves will be obtained (Table 1). *Kepler*, being a space mission, has no photometric errors due to imperfect weather conditions. The minimum quality of these light curves – depending on the ability to remove systematic errors – will approach or exceed 0.5% (Table 2), 5 – 10 times better than the typical ground-based light curves. This will enable a higher fraction rotation periods to be determined than the present ~25%. A fraction of objects, ~15%, will be binaries, for which masses can be determined from Kepler's laws.

This large database will be homogeneous in quality with accurate error bars. All observed objects will be included in the database regardless of whether a period could be determined or not, so that reporting bias will be eliminated. Repeat observations of a subset of objects (number TBD) will provide a data quality test.

A subset of ~250 of the brighter NEAs should yield 3-dimensional structures from tomography. The long uninterrupted data trains obtainable with Kepler will greatly assist in shape reconstruction.

Repeated observations a few years apart will measure, or put limits on, YORP spin-up/-down effects (e.g. Bottke et al. 2006).

Knowing the systematic characteristics of a large sample of NEAs will allow these properties to be applied to the larger population of NEAs, particularly if a LINNAEUS-like program operates simultaneously. This will illuminate their origins and histories. This understanding will then greatly alleviate the unknowns involved in selecting mission targets for robotic or human exploration, and in assessing the potential impact effects from PHOs, and deciding on deflection strategies.

## 2.10 ISSUES TO BE ADDRESSED IN A STUDY PHASE

Before *Kepler* can be used for this program a number of issues will need to be addressed:

1. Photometric accuracy across many pixels and, potentially, several CCD chips. Systematic errors are likely to dominate over Poisson errors. If they exceed ~2% then the accuracy advantage of *Kepler* will be minor. The advantage of having a large homogenous data set will remain.
2. Robust estimates of the number of NEAs expected within the *Kepler* field of regard per year above the threshold V magnitudes for:
    a. Period and aspect ratio determination,
    b. Detailed tomography.
3. Number of NEAs expected within the fields of other *Kepler* programs per year, above the threshold V magnitudes.
4. H magnitude (~size) distribution of the V<V(thresh) targets to roughly separate fast and slow rotators (Figure 3).
5. Proper motion rate distributions vs. H magnitude will give the time within the *Kepler* field of view.

6. Will the astrometry of the NEAs be adequate to position the apertures?
7. Conversely, could *Kepler* astrometry usefully improve the NEA orbit determinations?
8. Range of ToO warning times for new NEAs discovered during each year.
9. Policy on interrupts to other science programs.